\documentclass[a4paper]{jpconf}
\usepackage{graphicx}
\usepackage{color}
\usepackage{amsmath,amssymb,wasysym}
\usepackage{comment}

\begin{document}
\title{Primordial Gravitational Waves with LISA}

\author{Angelo Ricciardone}

\address{Faculty of Science and Technology, University of Stavanger, 4036, Stavanger, Norway}

\ead{\texttt{angelo.ricciardone@uis.no}}

\begin{abstract}
Primordial Gravitational Waves are the next target of modern cosmology. They represent a window on the early Universe and the only probe of the physics and microphysics of the inflationary period. 
When the production of GWs happens in scenarios richer than the standard single-field slow-roll, the GW signal becomes potentially detectable also on scales smaller than the Cosmic Microwave Background.~LISA will be extremely complementary to CMB experiments to extract information about primordial inflationary models and in particular to probe phases of the inflationary period for which we have very poor knowledges.
\end{abstract}

\section{Introduction}

100 years ago Einstein wrote the theory of General Relativity and Gravitational Waves (GWs) were a precise prediction of his theory. In the Einstein theory GWs are wave solutions of the linearized weak-field equations; more precisely they were defined as transverse waves of spatial strain that travel at speed of light, generated by time variations of the mass quadrupole moment of a source.\\
Many efforts during the last 100 years have been done to detect GWs, both through their direct or indirect effects. 
They are generated from cosmological and astrophysical sources and they represent a powerful tool to shed light on the physics responsible for their generation. In fact GWs of cosmological origin would allow for a clear understanding of the physics of the early Universe and they would give information on the quantum nature of gravity and fields; on the other hand GWs from astrophysical sources are important to test better General Relativity (GR) in the strong field limit and to study compact objects and their properties.

Concerning primordial GWs we can distinguish two main types of backgrounds: those expected from inflation, and those expected from mechanisms active between the end of inflation and the onset of Big Bang NucleoSynthesis (BBN). The first category includes the standard background of GWs due to tensor vacuum fluctuations $h_{ij}$ of the Friedman-Lemaitre-Robertson-Walker (FLRW) metric during the inflationary period as well as the ones generated by other mechanisms still active during inflation, like GWs generated by the presence of extra-field(s), or by the effects of different symmetry pattern during inflation. In the second category we can mention GW backgrounds expected from very different sources: GWs produced during (p)reheating~\cite{Khlebnikov:1997di,Easther:2006gt,GarciaBellido:2007dg,Dufaux:2007pt,Figueroa:2013vif,Figueroa:2014aya,Figueroa:2016ojl,Antusch:2016con}, or by phase transitions~\cite{Kosowsky:1991ua,Kamionkowski:1993fg,Caprini:2007xq,Caprini:2009fx,Caprini:2015zlo,Weir:2016tov} and cosmic defects~\cite{Vachaspati:1984gt,Figueroa:2012kw,Binetruy:2012ze,Sanidas:2012ee}, or even GWs coming from the merging of primordial black holes (PBHs)~\cite{GarciaBellido:1996qt,Clesse:2016ajp,Bird:2016dcv,Cholis:2016kqi}.
In this contribution we will describe the GW backgrounds expected from different inflationary models, and we briefly mention about GW backgrounds expected from post-inflationary phenomena and GW production in alternative scenarios of inflation.\\
In the standard model of inflation, where the accelerated expansion is driven by a slow-roll scalar field GWs are expected to be generated by the amplification of the tensor vacuum modes (metric fluctuations) and their spectrum amplitude, almost scale invariant, is directly connected with the energy scale of inflation. This GW backgound leaves a precise imprint on the Cosmic Microwave Background (CMB): a curl-like pattern in the polarization (B-modes). Up to now we do not have a detection of this primordial signal, but only an upper bound, from which we can infer that the spectral energy density at present time is $h_{0}^2\Omega_{\rm{GW}}\left(f\right)\approx 5 \cdot 10^{-16} (H/H_{\rm max})^2\,,$ where $H$ is the Hubble parameter during inflation and $H_{\rm max} \simeq 8.4\cdot 10^{13}$ GeV is the current upper bound on the energy scale of inflation~\cite{Ade:2015lrj}. This GW background represents an irreducible contribution expected from any inflationary scenario and its detection would help to differentiate inflationary models from each other.
Its tiny amplitude makes the detection extremely challenging both for CMB experiments and  for the current and planned direct detection experiments like LIGO or LISA, since the amplitude of the spectrum is well below their sensitivities. 

A way out to overcome this tiny amplitude is to consider inflationary scenarios richer than the standard single-field, for instance models where there are secondary fields with arbitrary spin~\cite{Cook:2011hg} coupled to the inflaton either directly  or only through a gravitational coupling~\cite{Biagetti:2014asa,Namba:2015gja,Shiraishi:2013kxa} or models where new symmetry patterns are allowed during inflation~\cite{Endlich:2012pz,Cannone:2014uqa,Bartolo:2015qvr,Ricciardone:2016lym}.\\
Depending on the model, i.e.~on the type of secondary field(s), the GW signal generated during inflation could be potentially interesting for the detectability by LISA. We can distinguish several cases, depending on the nature of the secondary field(s): (\textit{i}) In the case in which the inflaton is a pseudo-scalar (axion) coupled with a gauge field, the latter are excited upon the rolling of the axion during inflation. Towards the end of inflation, the tensor spectrum is enhanced for the corresponding modes leaving the Hubble radius during the last e-folds of inflation. This results in a blue-tilted chiral spectrum of GWs, with an over production of one of the tensor polarizations versus the other; (\textit{ii}) A spectator field during inflation, with a small speed of sound $c_s\ll 1$, and interacting only gravitationally with the inflaton field, sources second order tensor fluctuations and gives rise a blue tensor spectrum. (\textit{iii}) Considering non-Abelian $SU(2)$ vector fields coupled to the inflaton through the interaction $\chi F \tilde{F}$~\cite{Adshead:2012kp} can also bring to an enhancement of tensor modes and parity violating signal.
The time dependent background assumed during the inflationary period has suggested to assume that the inflaton is only time dependent during inflation. However considering the possibility to have (also) a space-dependent field during inflation can have potentially interesting features, in particular in the tensor sector.  When, besides the time-reparameterization invariance ($t\rightarrow t+\xi_{0}$), also the space-reparameterization invariance ($x_{i}\rightarrow x_{i}+\xi_{i}$) is broken during inflation, tensors can acquire a non-trivial mass changing the dispersion relation~\cite{Cannone:2014uqa}. This can enhance the spectrum of GWs on small scales becoming a potential interesting signal for interferometers, like LISA, and opening the possibility to test the symmetry patterns valid during the inflationary period at small scales.

A  production of GWs  is expected also during the post-inflationary period.  An interesting case is when, in scenario like mild-waterfall hybrid inflation, large peaks appear in the matter power spectrum that later collapse forming primordial black holes, at horizon reentry during the radiation era. These PBHs can merge generating a stochastic background of GWs potentially detectable by the current aLIGO detector and most probably by LISA.  Considering the period after the end of inflation, the so-called reheating phase, when the inflaton starts to oscillate and it couples with some ``preheat'' fields, there is the possibility, when the oscillations become large and coherent to generate a non-perturbative excitation of these fields and a consequent production of GWs. Typically the mechanisms through this happens are: i) parametric resonance~\cite{Kofman:1994rk} or ii) tachyonic preheating~\cite{Felder:2000hj}. This depends strongly on the given scenario (inflationary potential/set-up), and on the specific form and strength of the interaction between the inflaton and the preheat field. It also depends on the spin of the preheat field, whether it is a fermion or a boson. The non-perturbative excitation of the preheat field only occurs within specific bands of momenta, so this translates into the development of rapidly evolving large inhomogeneities of the energy density of the field. Hence, the large time-dependent gradients of the field provide an anisotropic-stress over the background, which naturally acts as a significant classical source of GW. The resulting spectrum of GW typically has a well defined peak, with an infrared slope decaying as $\propto k^3$, and an exponentially suppressed large-k tail. The position and height of the peak depend on the parameter space of each scenario. The amplitude of these backgrounds can be very large, but they naturally peaked at frequencies higher than the ones probed by interferometers like LISA.
In the category of production of GWs during the post inflationary period can be mentioned also the possibility to generate GWs during a  {\it stiff} phase after inflation, when the inflaton eventually behaves like a fluid  with an equation of state $w>1/3$~\cite{Spokoiny:1993kt}. Potentially, this could enhance the inflationary signal so much that it becomes detectable in the range of frequencies probed by LISA, while being at the same time compatible with the CMB constraints. However this scenario deserves further analysis.\\
Interestingly, gravitational wave generation can happen also in the context of string cosmology, in particular in the pre-big bang scenario. In~\cite{Gasperini:2016gre} has been shown that a generation of stochastic background of GWs with amplitude and frequency in the range probed by interferometers like aLIGO/Virgo and/or eLISA is possible, for a wide range of the parameter space. And in particular, this is consistent with the constraints coming from CMB about temperature anisotropies.\\
The mentioned scenarios are only a partial list of models where GWs produced in the early Universe are potentially detectable by an experiment like LISA, see i.e.~\cite{Guzzetti:2016mkm} for a recent review. This represents a way to shed light on the early Universe and in particular to test the latest stage of inflation, or the presence of other field besides the inflaton during inflation, or to understand possible symmetry breaking pattern during inflation, or to study the post inflationary phase for which we have poor information or finally to confirm or disprove pre-big bang models or to impose new significant constraints on the related cosmological parameters.

\section{Detection of GW}
After the recent detection by Advanced LIGO of (almost) three GW signals generated by the merging of two massive BHs~\cite{Abbott:2016blz,Abbott:2016nmj}, the detection of primordial GWs has become a feasible target of modern cosmology. Due to the weakness of the gravitational interaction, GWs cannot interact with the surrounding medium, and hence immediately after production they decouple and travel freely through the Universe until today. While CMB photons decouple at energy around $0.3$ {\rm{eV}}, primordial GWs are expected to decouple at very high energy scale (around the Planck scale). Then this would open a window on the very early Universe. Actually there are two main ways to try to detect GWs: through their indirect effect on the CMB and LSS or through their direct effect (of displacement) on the arms of interferometers. Let's briefly discuss about these ways, and report the most recent limits about the GW amplitude from different experiments:

{\it{Indirect detection}} The influence of GWs when photons decouple
from ordinary matter forming the Last Scattering Surface (LSS) leaves a precise imprint: in particular the CMB radiation becomes linearly polarized showing a curl-like� (B-mode) pattern. The amplitude of this signal is extremely small since only few CMB photons get polarized, then its detection is rather hard. The main contribution from primordial GWs is expected to be in the BB power spectrum mostly at very large scales ($\ell < 150$), but astrophysical foregrounds, galactic dust and gravitational lensing contaminate this signal making its detection even harder.   Current data actually provide only an upper bound on the amplitude of the tensor power spectrum defined through the tensor-to-scalar ratio $r\equiv \mathcal{P}_{T}/\mathcal{P}_{S}$, where $\mathcal{P}_{T}$ and $\mathcal{P}_{S}$ are the amplitudes of the tensor and scalar power spectra respectively, evaluated at a reference scale.  The joint analysis by BICEP2, Keck Array data, Planck polarization and WMAP9 23 GHZ and 33 GHZ maps gives, at  a pivot reference scale $k_{*}=0.05$
${\rm{Mpc^{-1}}}$, a bound on $r_{0.05}<0.09$ at $95\%$ C.L.~\cite{Array:2015xqh},  assuming a scale-invariant GW power spectrum. Assuming the consistency relation between the tensor-to-scalar ratio  and the tensor spectral tilt $r=-8n_{T}$, valid for single-field slow-roll models of inflation, the limit becomes tighter: $r_{0.05}<0.07$ at $95\%$ C.L. 
There are several current or forthcoming experiments devoted to B-mode detection: among the space-borne we can mention PRISM, COrE or PIXIE, while among the ground-based we can mention BICEP2, SPTPol, ACTPol, Polarbear, Spider. See the recent~\cite{Abazajian:2016yjj} for the next generation of ground-based CMB projects. 	 

{\it {Direct detection}} Direct detection GW experiments are based on laser interferometry and they essentially account for the effect of distorsion that a passing GW can have on the length of the arms of the interferometer. They are mainly devoted to the analysis of GWs produced by compact binary systems or mergers of massive black holes, neutron stars or extreme-mass-ratio inspirals. However, as seen in a series of papers of the LISA collaboration, LISA will have the capabilities to detect or at least to constrain, not only astrophysical sources, but also cosmological sources, like GWs produced during first order phase transition, from cosmic defects~\cite{Caprini:2015zlo} and probe the present expansion of the Universe~\cite{Tamanini:2016zlh}.  Actual gravitational wave detectors already in operation on the Earth are LIGO, Virgo and GEO600. They are designed to be sensible to relative changes in the length of the arms of the order of $\Delta L/L \simeq 10^{-21}$. LIGO and Virgo have arm lengths of 4 and 3 Km respectively while GEO600 of about 600m. They were built to be sensitive in the range of frequency between few Hz and 10 kHz and they are mainly devoted to the detection of GWs generated by astrophysical objects. Their sensitivity however is limited by seismic noise and by practical reason as the length of the arm, so this pushed the community to propose the first gravitational wave observatory in space. LISA represents the main candidate for this mission~\cite{AmaroSeoane:2012km}. The LISA concept, still based on laser interferometry, will be composed by a constellation of three spacecrafts, arranged in an equilateral triangle with million-kilometre arms (5 million km or 2 million km) flying along an Earth-like heliocentric orbit with a duration of the mission between 2 and 5 years. It will have the main sensitivity in the frequency domain between mHz and Hz and it will be extremely complementary to Earth based interferometers~\cite{Sesana:2016ljz}.\\
Other forthcoming or planned gravitational wave detectors are KAGRA~\cite{Somiya:2011np}, LIGO-India~\cite{LIGO-India} and Einstein Telescope~\cite{Sathyaprakash:2012jk} that will be also mainly devoted to study astrophysical sources of GWs. In order to have GW interferometers sensible to a primordial GW signal generated in a standard inflationary scenario, then we have to refer to more futuristic experiments like BBO and DECIGO~\cite{Kawamura:2011zz}.\\
Nowadays, from a non-detection of primordial signal from the aformentioned detectors we have current and forecasted upper bounds on the GW spectral energy density:  the LIGO/Virgo collaboration provided an upper limit $\Omega_{GW}< 5.6\times 10^{-6}$ at $95\%$ C.L. for $f\in(41.4,169.25)$~cite{}. The upgraded aLIGO detector is expected to lower this bound to $\Omega_{GW}< 10^{-9}$~\cite{TheLIGOScientific:2016wyq}, while LISA (best configuration) is expected to reach  $\Omega_{GW}\lesssim 10^{-14}$~\cite{antoine}. 
   
{\it{Effect on LSS}} Primordial GWs can leave their imprint also on Large Scale Structures (LSS), in particular a primordial long-wavelength tensor mode may leave its imprint on the primordial scalar power spectrum and consequently on the matter power spectrum~\cite{Dimastrogiovanni:2014ina,Creminelli:2014wna}. The typical effects are the generation of a local power quadrupole in the matter power spectrum, that gives rise to an apparent local departure from statistical isotropy, and  a non-Gaussian four-point function. These effects, incorporated in the $\langle\gamma \zeta\zeta\rangle$ (tensor-scalar-scalar) correlation function, have observational consequences, even if the tensor perturbations cannot be detected  directly. In particular the quadrupole anisotropy is generated by a primordial GW with wavelength larger that the size of the galaxy survey while the non-Gaussian effects may arise from GWs of wavelength comparable to the survey size. These observables are useful quantities to discriminate among the plethora of inflationary models and the next galaxy surveys will have the possibility to probe this {\it{indirect}} GW effect~\cite{}. 
From this brief description of the three main strategies to detect GWs it becomes clear the importance of the complementary among the different experiments in order to probe different (energy) scales and automatically to test several stages of the primordial Universe, as clearly explained in Table 1.
\begin{table}[t!]\footnotesize
\centering
\begin{tabular}{|l|c|c|}  \hline
 & $k \;\; \left[ {\rm Mpc}^{-1} \right] $ & $ N_{\rm estim.} $\\  [-1ex]\hline 
CMB / LSS & $10^{-4} - 10^{-1}$ & $56-63$ \\ [-1ex] \hline
$y-$ \& $\mu-$distortions & $10^{-1} - 10^4 $ & $45-56$ \\  [-1ex]\hline
$P_\zeta \,\rightarrow\, {\rm PBH} \, \rightarrow \, {\rm GW}$ @ LISA  & $10^5 - 10^7 $ & $38-41$ \\  [-1ex] \hline
$P_\zeta \,\rightarrow\, {\rm PBH} \, \rightarrow \, {\rm GW}$ @ AdvLIGO  & $ 10^7 - 10^8 $ & $35-37$ \\  [-1ex]\hline
$P_{\delta g}  \, \rightarrow \, {\rm GW}$ @ LISA  & $10^{11} - 10^{14}  $ & $22-28$ \\  [-1ex]\hline
$P_{\delta g}  \, \rightarrow \, {\rm GW}$ @ AdvLIGO  & $10^{16} - 10^{17}  $ & $15-17$ \\ \hline
\end{tabular}
\caption {List of observational windows on inflation in correspondence of the wavenumber of the primordial modes  and estimated number of efolds at which those modes exited the horizon.  Taken from~\cite{Garcia-Bellido:2016dkw}.} 
\label{tab:window}
\end{table}
\section{Inflationary consistency relation}
\begin{figure}[t]
\begin{center}
\includegraphics[width=5.8cm]{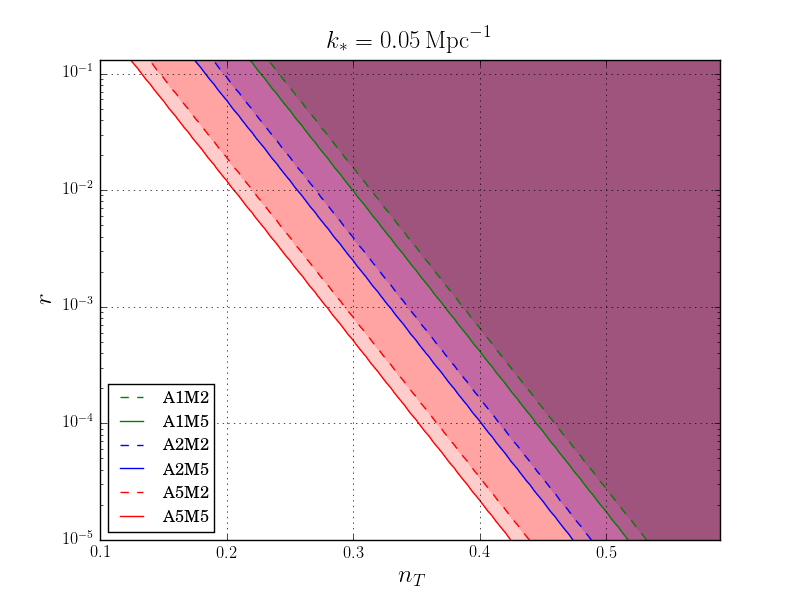}
\includegraphics[width=9.5cm]{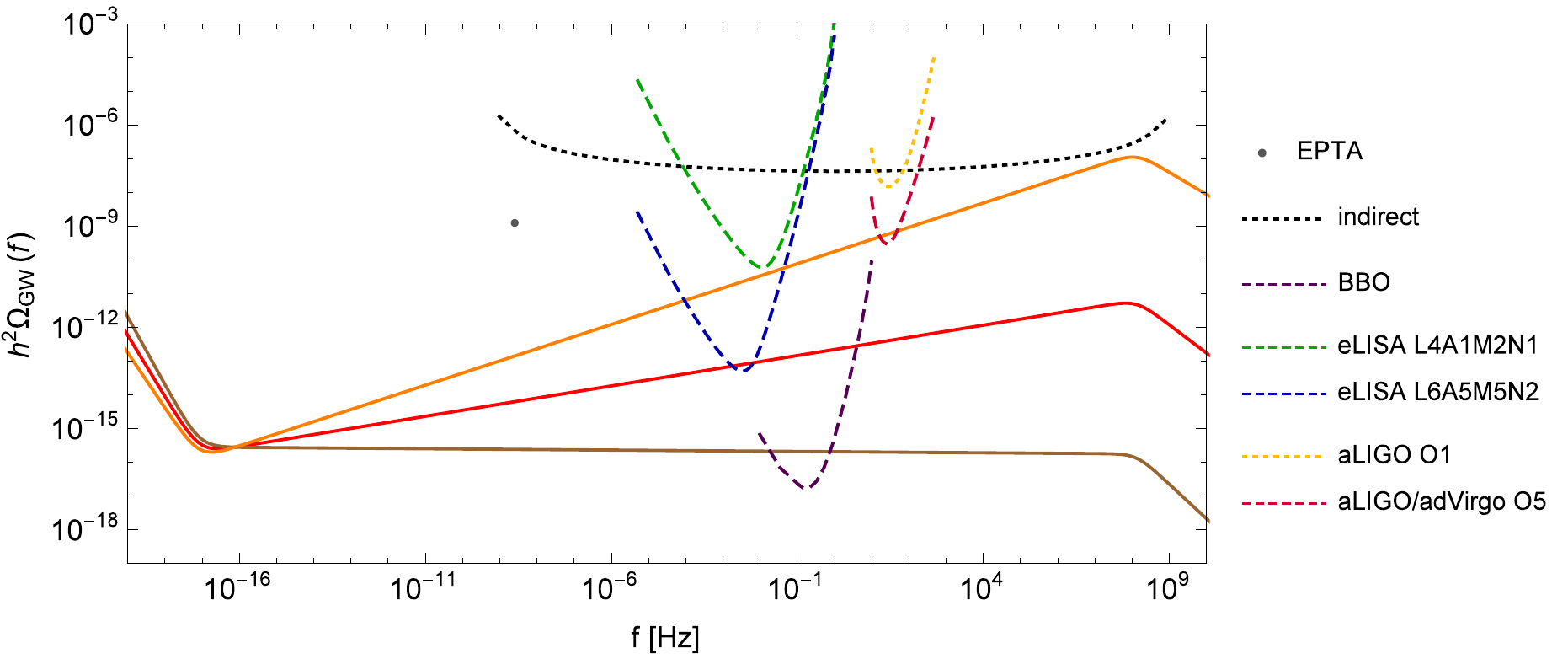}
\end{center}
\caption{Left: limits on the tensor spectral tilt $n_T$ and the tensor-to-scalar ratio $r$ for six LISA configurations. Taken from~\cite{Bartolo:2016ami}. Right: GW energy spectrum for different value of the tensor spectral tilt vs sensitivities of different experiments (orange $n_{T}=0.36$, red $n_{T}=0.18$, brown $n_{T}=-r/8$). Taken from~\cite{Guzzetti:2016mkm}; see also~\cite{Lasky:2015lej}. }
\label{fig:rvsnt}
\end{figure}

The majority of the models that produce a GW signal during the inflationary period are characterized by a primordial power spectrum with a power law shape, that can be directly connected to the present spectral GW energy density. This possibility to parametrize the GW energy-density spectrum by a power law opens the possibility to constrain cosmological parameters which are strictly connected with the inflationary period. In particular, we can constrain   
the GW spectral index $n_{T}$ and tensor-to-scalar ratio $r$. Then assuming a power law spectrum like
$\Omega_{GW}(f)=\Omega_{GW}^{CMB}\left(\frac{f}{f_{CMB}}\right)^{n_T}$, 
we can re-express $\Omega_{GW}^{CMB}$ in terms of the tensor-to-scalar ratio {\it{r}} and the amplitude of the primordial scalar power spectrum at CMB scales for which we have precise limits, given by Planck \cite{Ade:2015lrj}. In this way we can combine constraints on {\it{r}} and $n_{T}$ from the CMB scale with constraints to $\Omega_{GW}(f)$ and $n_{T}$ from direct detection experiments.\\
Up to now a constraint on $n_{T}$ have been obtained combining {BICEP2/Keck Array and Planck (BKP), Planck $2013$, WMAP low $\ell$ polarization, HST data, BAO measurements from SDSS
and the upper limit on the energy density of stochastic GW background from
LIGO: $n_{T}=0.06^{+0.63}_{-0.89}$ at $95\%$
C.L. \cite{Meerburg:2015zua}, in correspondence of a best fit for the tensor-to-scalar ratio of $r_{0.01}=0.02$.} \\
Recently, it has been shown how CMB experiments alone are not able to put strong constraints on the spectral tilt ($n_{T}\lesssim5$ at $95\%$
C.L. for $r_{0.01}=0.02$  \cite{Lasky:2015lej}), even in the case of a detection of B-modes, since they focus on a narrow range of frequency around $10^{-16}${\rm{Hz}}; so, it becomes clear the importance of the combination of several experiments that cover different range of scales.
In \cite{Lasky:2015lej, Cabass:2015jwe}  it has been pointed out how the combination of GW experiments on a large range of frequencies, as ground and space-based interferometers, indirect measurements and CMB, BAO, BBN, puts tighter constraints on the tensor spectral index. \\
As forecasted in~\cite{Bartolo:2016ami}, LISA  will have the ability to constrain better the spectral tilt $n_{T}$, considering $r$ consistent with the current CMB upper bound. The results are shown in Fig.~\ref{fig:rvsnt}, where are shown different LISA sensitivities; assuming the best LISA configuration with six links, five million km arm length and 5 year mission, the constraint for the spectral index turns out to be $n_{T}\lesssim 0.2$.This ability of LISA in constraining the tensor spectral tilt, can be seen as a test of the so-called consistency relation $r=-8n_{T}$ for primordial inflationary models. In fact, we know that, all single-field slow-roll models of inflation predict such a relation. Any evidence of violation of such a relation would be an indication of deviation from single-field slow-roll models.

\section{Gravitational Waves from Inflation}

\subsection{The standard inflationary GW background}

Gravitational Waves are with no doubt the smoking gun of an inflationary period. After the several predictions verified by the Planck satellite about a period of accelerated expansion~\cite{Ade:2015ava}, GWs represent the key test to finally confirm inflation. They are ubiquitous in all the inflationary models and are represented by the tensor perturbation $h_{ij}$ of the FLRW metric $g_{ij}=a^2(\delta_{ij}+h_{ij})$. They carry two physical degrees of freedom represented by the two polarizations and this is mathematically given by the trasverse-traceless conditions on the tensor perturbation ($\partial_{i}h^{i}_{\,j} = 0$, $h^{i}_{i}=0$). Their dynamics is described by a wave equation in an expanding background. 
These vacuum metric fluctuations are expected to be amplified during the inflationary period, when small-scale fluctuations are stretched to superhorizon scales. Being massless (or very light) quantum fields in an expanding background, they fluctuate with a magnitude proportional to their stored energy density, which through Einstein equations, is related to the (square) of the Hubble parameter during inflation. This fixes also the amplitude of the GW power spectrum and shows the importance of a detection of primordial GW signal.
In fact in the standard inflation scenario, where the accelerated expansion is driven by a slowly rolling scalar field, the inflaton, the amplitude of the primordial GW signal, expressed by the tensor-to-scalar ratio $r$, defines two important relations: $V^{1/4}\simeq10^{16} {\rm{GeV}} \left(\frac{r}{0.1}\right)^{1/4}$, which clarifies the relation between the energy scale of inflation and the amplitude of the GW spectrum, and $\Delta\phi/M_{Pl} \gtrsim 1.06 \left(\frac{r}{0.1}\right)^{1/2}$  which connects the amplitude of the tensor spectrum with the excursion of the inflaton during inflation. These two relations are a clear implication of the importance of the detection of  inflationary GWs since they would represent a power way to discriminate among inflationary models, to fix the energy scale regime of the inflationary period and related issues about quantum gravity effects.\\
In single-field scenarios such a signal is expected to have a small amplitude and to peak at large scales where the level of foreground contamination is quite large, that makes it detection quite challenging. Nowadays we do not have a detection of such a primordial signal but only current limits as seen before. Considering this bound and the almost scale-invariance of the spectrum, predicted by the simplest models of inflation involving a slow rolling scalar field, constrains this GW signal in a range of frequencies and amplitude out of current direct GW detectors and also from the LISA capabilities. 
\subsection{Beyond the standard inflationary GW background}
\begin{figure}[t]
 \begin{center}
\includegraphics[width=7.2cm]{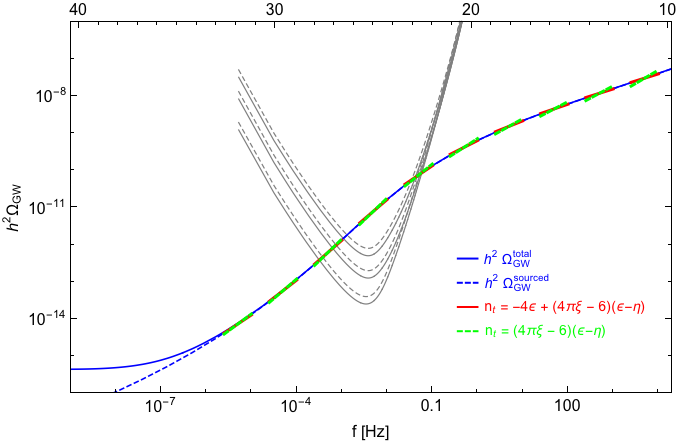}
\includegraphics[width=0.24\textwidth]{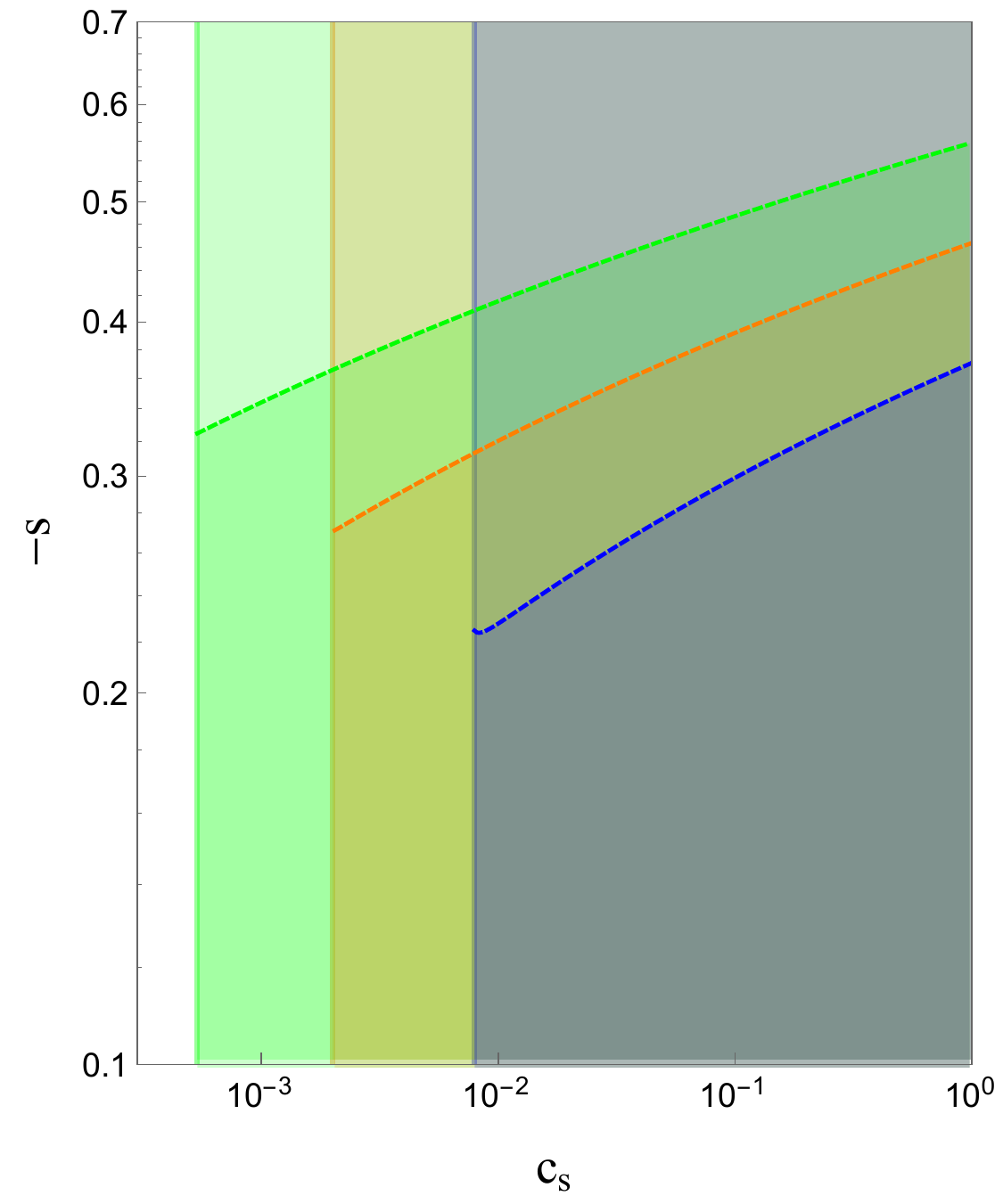}
\includegraphics[width=0.24\textwidth]{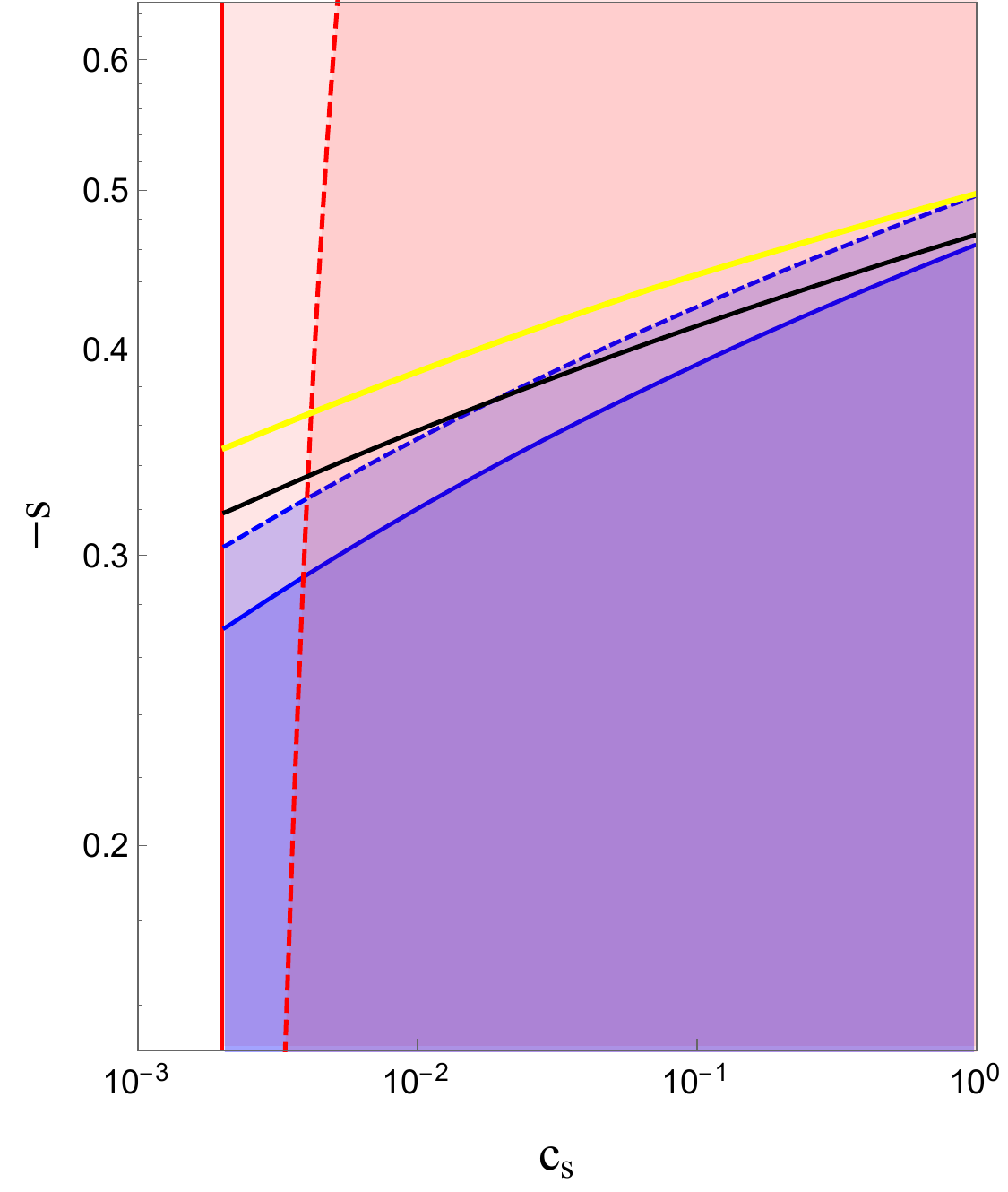}
\end{center}
    \caption{Spectrum of GWs in particle production model (left panel) and parameter space that LISA can probe with different configurations for the spectator field scenario (right panel). Taken from~\cite{Bartolo:2016ami}.}
  \label{bbnligo}
\end{figure}

\begin{figure}[t!]
\begin{center}
\includegraphics[width=5.0cm]{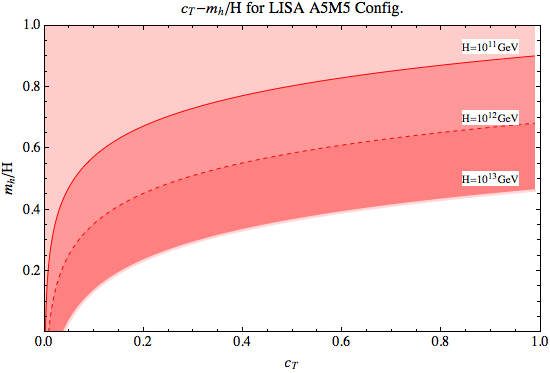}
\includegraphics[width=7.0cm]{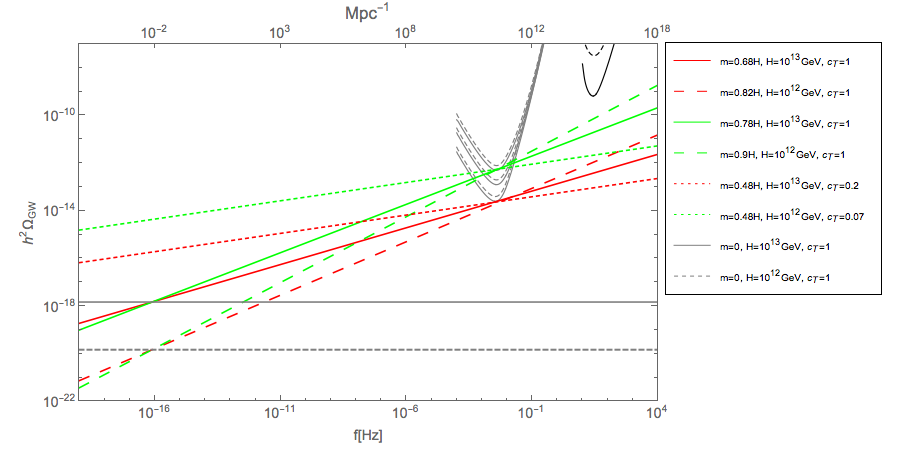}
\end{center}
\caption{Parameter space that LISA can probe with the ``best'' (A5M5) configuration (left panel)  and spectrum of GWs in the EFT framework (right panel). Taken from~\cite{Bartolo:2016ami}.}
\label{fig:massspeedt}
\end{figure}

As soon as the inflationary scenario becomes richer than standard single-field slow-roll inflation the GW predictions can change. In particular when there are additional fields during inflation or when the pattern of symmetry is modified with respect to the usual time-dependent background, the GW signal can become large and blue tilted, becoming a possible target for detection at scales of interferometers like LISA and/or aLIGO.\\
Whenever secondary field are present during inflation, coupled directly or only gravitationally to the inflaton, the wave equation for the tensor gets modified by a source term,  which generates an extra ``sourced" contribution which adds to the irreducible background signal given by vacuum metric fluctuations. In particular the equation governing the GW evolution becomes		
$\ddot{h}_{ij}\left(\bold{k},t\right)+3H\,\dot{h}_{ij}\left(\bold{k},t\right)+k^2\, h_{ij}\left(\bold{k},t\right)=\frac{2}{M_{{Pl}}^2} \Pi_{ij}^{TT}\left(\bold{k},t\right)$,
where a dot denotes derivative with respect to {\it{t}}, {\it{H}} is the Hubble rate, ${\it{M_{Pl}}}$ is the reduced Planck mass, {\it{k}} is the physical momentum, and $\Pi_{ij}^{TT}$ is the source of  GWs, corresponding to the transverse-traceless part of the anisotropic stress $\Pi_{ij}$. The latter is directly connected to the nature of the additional sources, i.e.~if they are scalars or vectors. In the following models we will see its specific form. In many cases this sourced contribution can be larger that the vacuum contribution, generating a GW signal large at small scales and still compatible with CMB bounds. \\
{\textbf{Particle production during Inflation}} Among the different mechanisms of particle production during inflation and subsequent enhancement of the GW signal potentially detectable by LISA interferometer, still compatible with the constraints from the CMB, the models involving $U(1)$ gauge field, coupled with a pseudo-scalar inflaton field, look the more promising.
These models solve also a key problem of inflation that is the protection of the flatness of the inflaton potential from radiative corrections. In fact the protection is naturally present since the pseudo-scalar axion respects an approximate shift symmetry.\\
In the simple model \cite{Barnaby:2011qe} the inflaton {$\phi$} is a pseudo-scalar field and it is coupled with a gauge field {$A_\mu$} with the relevant Lagrangian operator and the source tensor given by 
$\mathcal{L}  \supset  - \frac{{\phi}}{4f}{} F_{{\mu\nu}}\tilde{F}^{{\mu\nu}}\;\;\;\;\;\;	\Rightarrow \;\;\;\;\;\; \Pi_{ij}^{TT}\propto (E_{i}E_{j}+B_{i}B_{j})^{TT}$
where $\textit{f}$\ is the axion decay constant,
$F_{\mu\nu}\equiv \partial_{\mu}A_{\nu}-\partial_{\nu}A_{\mu}$ is the field strength of the gauge field $A_{\mu}$ and $\tilde{F}_{\mu\nu}$ is its dual; $\mathbf{E}=-\mathbf{A'}/a^2$, $\mathbf{B}=\nabla\times\mathbf{A}/a^2$ and the source term arises not from the $F\tilde{F}$ but from the term $FF$ also present in this kind of models. The dynamics of the scenario is the following: the (slowly-rolling) inflaton excites, through its coupling, one helicity of the vector field, that in turn sources the GW of one definite chirality in an interaction like $\delta A + \delta A \rightarrow \delta g$. However also the inflaton perturbations are sourced through the inverse decay of the gauge field $\delta A + \delta A \rightarrow \delta {\phi}$. This suggests a high level of scalar non-Gaussianity that, being strongly constrained by the non-observation in the CMB, puts a limit on the parameter $\xi$~\cite{Ade:2015ava}. In fact the production is exponentially proportional to the parameter
$\xi\equiv \frac{\dot{{\phi}}}{2 f H}$, 
that sets the exponential amplification of the gauge mode $(A_{+}\propto e^{\pi \xi})$ and is constrained by observations from CMB and LSS to be smaller than $2.5$ at $95\%$ C.L. in order not to generate too much non-Gaussianity. Such small value suppresses the amplitude of the GW signal, however such a quantity $\xi$ is depending on the time derivative of the inflaton that towards the end of inflation speeds up. Therefore, in  this model we naturally expect a growth of $\xi$, and a consequent exponential increase of GW production, to at level that may  be probed at LISA and at terrestrial interferometer scales \cite{Cook:2011hg, Crowder:2012ik, Barnaby:2011qe, Domcke:2016bkh}, like shown in Fig.~\ref{bbnligo}. A detailed analysis of this model in connection with a GW signal potentially detectable at LISA scales has been carried out in~\cite{Bartolo:2016ami}. In particular they used a local approach, which allows to study the phenomenology of the model within a well-defined observational range, and a global approach which cover a full observational range. Both approaches highlight the ability of the LISA detector to extract relevant information about the parameters of the model giving in this way the possibility to test stage of inflation that are far from the scales probed by CMB. In particular the analysis shows the complementary between CMB observables, coming from power spectrum and bispectrum analysis and parameters related to the model and connected to the GW spectrum.
As anticipated, this scenario shows peculiar features that are extremely useful as observational test: in particular a GW signal  that is completely chiral~\cite{Sorbo:2011rz}, which means that one of the polarizations of the gravitational field shows a larger spectrum amplitude, and a highly tensor non-Gaussian signal with a precise relation between the bispectrum and the power spectrum: $k^6 \, \langle h^3 \rangle_{\rm equil}' \simeq 23 P_{\rm GW}^{3/2} $ . If detected, these properties would be a smoking gun for the cosmological origin of this GW signal and its nature. \\
{\textbf{GWs from spectator field during inflation}} Also the presence of other fields, not directly coupled to the inflaton, during inflation can generate interesting features in the GW spectrum. In particular considering a {\it{spectator}} field with a small speed of sound during inflation, where a spectator is usually referred to a field that does not influence the background inflaton dynamics, we can generate a blue-tilted primordial power spectrum for tensor modes, allowing to a possible detection of this signal at LISA scales. The presence of a spectator field, however, is expected to source also scalar fluctuations, as it happens in particle production by gauge field as seen before, and these are well constrained by current CMB measurements~\cite{Ade:2015lrj}. This mechanism is able to generate a small value of the tensor-to-scalar ratio on CMB scales, as shown in a particular realization~\cite{Fujita:2014oba}, while can instead produce more tensor power on smaller scales thanks to the presence of a {\it{source}} term in the equation of motion of the tensor, given by the spectator.\\
Considering the model introduced in~\cite{Fujita:2014oba} and analyzed also in~\cite{Bartolo:2016ami} with
$\mathcal{L}  \supset P\left(X,\sigma\right)\;\;\;\;\;\;	\Rightarrow \;\;\;\;\;\; \Pi_{ij}^{TT}\propto P_{X}(\partial_{i}	\delta \sigma \partial_{j}\delta\sigma)^{TT}$,
where $\sigma$ is the spectator field, $X=\frac{1}{2}\partial_{\mu}\sigma\partial^{\mu}\sigma$ and $P$ is a generic function of $X$ and $\sigma$, an interesting case appears when the spectator sound speed  $c_{s}\equiv P_{X}/\left(P_{X}+P_{XX}\dot{\sigma}_{0}^{2}\right)$ (where $\sigma_{0}$ is the background value) and its time variation $s\equiv \dot{c}_{s}/H c_{s}\neq 0$ are very small, in particular when  $c_{s}\ll 1$ and $\left|s\right|<1$. Both in the scalar and tensor sectors the power spectra will be the sum of the usual vacuum contribution and of the sourced contribution generated by the presence of the spectator field perturbation. Particular regimes for the parameter  $c_s$ and $s$ allow to obtain a sourced GW contribution with a sufficiently large amplitude, in principle detectable by LISA, and at the same time keeping a small amplitude at CMB scales. The parameter space of the model is characterized by the Hubble rate during inflation, the speed of sound of the spectator $c_s$ and its time variation $s$, since: $P_{S}\equiv P_{S}(\epsilon, H, c_{s})$, $P_{GW}\equiv P_{GW}(H, c_{s})$ and the spectral tilts $n_{T}^{\left(\sigma\right)}=-4\epsilon-3s$, $n_{S}^{\left(\sigma\right)}-1=-4\epsilon-7s$. So taking into account the actual limits on $\epsilon$ from CMB ($\epsilon<0.0068$ at $95\%$ C.L.~\cite{Ade:2015lrj}) and on the amplitude of the scalar perturbation ($A_{0.05}=2.21\cdot 10^{-9}$ at $68\%$ C.L.~\cite{Ade:2013zuv}), it is possible to scan the parameter space $s-c_s$ probed by the six-different LISA configurations for different values of the Hubble rate during inflation. Fig.~\ref{bbnligo} reports some of the results found in~\cite{Bartolo:2016ami},  showing the ability of different LISA configurations to scan the parameter space of the model. The region below each curve represents the parameter space left open by a non-detection of a GW signals associated to the spectator scenario. 
Bounds for this scenario come also from other current observations like Big Bang Nucleosynthesis, or from the aLIGO O1 which both provide upper bound on the spectral energy density of GW. These bounds are also reported in the right panel of Fig.~\ref{bbnligo} and is possible to compare how much the LISA detector will eventually improve these current bounds. \\
As a consequence of the increased power in the tensor sector due to the spectator field, also the scalar sector, and in particular the sourced contribution is expected to increase at small scales. However on such small scales we have to be consistent with bounds coming from PBHs. This bound would reduce drastically the parameter space accessible to LISA, however the modellization of PBH physics has many uncertaintes which allow to still consider the future constraints by LISA as the more robust. Whenever this modellization of the behaviour of PBH will become more solid, then LISA will still be a validation test for the modellization. An interesting perspective on how much PBH constraints limit the ability of observation of GWs by LISA is addressed in~\cite{Garcia-Bellido:2016dkw}.\\
{\textbf{GWs in the Effective Field Theory of space-reparameterization}} There are also interesting case of large GW signal at small scales without invoking the presence of an extra field. In particular, this can be realized in models where inflationary fields acquire vacuum expectation values along space-like directions~\cite{Endlich:2012pz, Bartolo:2013msa}. While time translation invariance ($t\rightarrow t+\xi_0$) is surely broken by the cosmological expansion during inflation, it is interesting to explore the possibility that space-reparameterization ($x_{i}\rightarrow x_{i}+\xi_{i}$) are also broken during the inflationary era. One writes
the most general set of operators that satisfy the symmetry requirements, and connects the coefficients of such operators with observable quantities. The effective field theory of inflation (EFTI) provides a powerful tool 
 for obtaining model independent  predictions of large classes of inflationary scenarios. It  requires  only information
 about the symmetries broken during the inflationary era, and on the number and nature of fields that drive inflation. 
The advantage of
EFTI is that  one has not to commit on specific realizations in order to deduce  observable consequences.
 If space-reparameterizations are broken, the graviton can acquire a mass and a non-trivial sound
speed during inflation~\cite{Bartolo:2015qvr}. At the quadratic level in perturbations, in an EFT approach, the action for graviton fluctuations $h_{ij}$ around a conformally flat
FLRW background can be expressed
as in~\cite{Cannone:2014uqa, Bartolo:2015qvr,Ricciardone:2016lym}:
%
$\mathcal{L}_{h}=\frac{M_{Pl}^2}{8}\,\left[\dot h_{ij}^2 -\frac{c_T^2(t)}{a^2} \,\left( \partial_l h_{ij}\right)^2
-m_h^2(t)\,h_{ij}^2\right]$
%
Using this action, the tensor power spectrum and its related spectral tilt are found to be:
${\cal P}_T\,=\,\frac{2\,H^2}{\pi^2\,M_{Pl}^2\,c_T^3}\,\left( \frac{k}{k_*}\right)^{n_T}$ and the spectral tilt $n_T\,=\,-2 \epsilon+\frac{2}{3}
\frac{m_{h}^2}{H^2}\left(1+\frac43\,\epsilon\right)$.
Hence, if the quantity $m_{h}/H$ is sufficiently large, one can 
get a blue tensor spectrum with no need to violate any energy condition in the early Universe ( null energy condition). This is a feature common in scenarios where space reparameterization is broken during inflation. In~\cite{Bartolo:2016ami} also this scenario has been analyzed finding interesting results. In Fig.~\ref{fig:massspeedt} we report the spectral GW energy density  computed starting from the power spectrum. This shows how the possibility to have a non trivial mass and sound speed for the graviton during inflation can enhance the tensor signal on small scales becoming a possible target for the LISA mission. In Fig.~\ref{fig:massspeedt} we report the perspective of LISA in constraining the parameter space $(c_{T} - \,m_{h}/H)$ for the best configuration of LISA: A5M5, with 5 million arm length, 5 year mission (left panel) . This analysis highligths the ability of LISA to probe different range of masses for the graviton, giving the possibility to test pattern of symmetries valid during the inflationary period.

\section{Beyond the power spectrum: parity violation, non-Gaussianity and extra-polarization states}

As seen in the previous sections a peculiar signature of some primordial GW signals is the chirality, then it is natural to ask if LISA will be able to detect parity violating GW signal. In particular we have seen that the breaking of parity invariance is a peculiar feature of models including a coupling between an axion-like field and gauge field(s). This has as a consequence that the spectra of the left and right-handed tensor modes generated by the amplification of the gauge field are different. This is a distinctive signature that could distinguish GWs of cosmological origin from astrophysical ones and also the GWs produced by these mechanism{s} from the vacuum ones. 
Parity violation signal can arise also in sectors of string theory ({\sl i.e.} type IIB, I, heterotic), in loop quantum gravity, leading to the question of whether the gravitational interaction is parity invariant in the strong field.
Prospects of measuring parity violation from the early Universe using gravitational terrestrial wave detectors have been done in~\cite{Crowder:2012ik, Seto:2007tn}.\\
At the same level of scalar also tensor non-Gaussianity represents  a powerful tool in order to help in constraining inflationary physics, see e.g.~\cite{Bartolo:2004if}. From CMB data we already have strict bounds on different configurations of the scalar bispectrum, and also (weaker) constraints on the tensor three-point function~\cite{Ade:2015ava}. Being in many models the GW signal produced by non-linear sources it is natural to ask if a detector like LISA will be able to measure the GW three-point function. As we have seen in the axion-gauge  model a large tensor bispectrum can be generated and the amplitude can provide upper bounds on the model parameter $\xi$, complementary to those coming from scalar CMB bispectrum~\cite{Cook:2013xea, Shiraishi:2013kxa}. Moreover in that specific scenario a precise relation between the power spectrum and the bispectrum is predicted which represents a smoking gun for such kind of GW signals.\\
Finally, as last open question, there is the possibility to consider additional GW polarizations. In fact, in alternative theories of gravity, GWs could have up to four additional (apart the ``plus" and ``cross" polarization models within standard General Relativity)  polarization modes. Then it is natural to ask if LISA will have the capabilities to detect extra GW polarization states.

\section{Conclusions}
Primordial GWs are one of the next targets of modern cosmology since they represent a unique opportunity to shed light on the physics of the early Universe and in particular to probe the microphysics of inflation. An irreducible GW background is an ubiquitous prediction of all the inflationary models and this represents a smoking gun of the primordial accelerated expansion. The primary probe of primordial GW is the polarization of the CMB and in particular the curl-free polarization pattern (B-modes). When the inflationary scenario is enriched by secondary fields, besides the inflaton, or by assuming new symmetry pattern, GWs become a target also on scales smaller than CMB, and in particular for interferometers like LISA. While the GW by vacuum fluctuations is not visible by the next generation interferometers, many well-motivated inflationary scenarios produce a signal that can be visible or that can be extremely useful to reduce the parameter space of such scenarios. It becomes clear that a detection of GW signal on the small scales accessible to LISA, will offer a window on scales of inflation on which we currently have little knowledge and will become of fundamental importance in order to provide constraints on tensor perturbations complementary to the CMB.

\section*{References}
\bibliographystyle{utcaps.bst}
\bibliography{LISAinflation}

\end{document}